\documentclass[aps,prd, final]{revtex4}
\usepackage[latin1]{inputenc}
\usepackage{amsmath}
\usepackage{amsfonts}
\usepackage{amssymb}
\usepackage{amsthm}

\ifx\pdftexversion\undefined
 \usepackage[dvips]{graphics}
\fi
\usepackage[pdftex]{graphicx}

\theoremstyle{remark}

\theoremstyle{definition}

\begin{document}

\title{The Local Effects of Cosmological Variations in Physical 'Constants'
and Scalar Fields \\
II. Quasi-Spherical Spacetimes}
\author{Douglas J. Shaw}
\affiliation{DAMTP, Centre for Mathematical Sciences, University of Cambridge,
Wilberforce Road, Cambridge CB3 0WA, UK}
\author{John D. Barrow}
\affiliation{DAMTP, Centre for Mathematical Sciences, University of Cambridge,
Wilberforce Road, Cambridge CB3 0WA, UK}
\date{\today}

\begin{abstract}
We investigate the conditions under which cosmological variations in
physical `constants' and scalar fields are detectable on the surface of
local gravitationally-bound systems, such as planets, in non-spherically
symmetric background spacetimes. The method of matched asymptotic expansions
is used to deal with the large range of length scales that appear in the
problem. We derive a sufficient condition for the local time variation of
the scalar fields driving variations in 'constants' to track their
large-scale cosmological variation and show that this is consistent with our
earlier conjecture derived from the spherically symmetric problem. We
perform our analysis with spacetime backgrounds that are of Szekeres-Szafron
type. They are approximately Schwarzschild in some locality and free of
gravitational waves everywhere. At large distances, we assume that the
spacetime matches smoothly onto a Friedmann background universe. We conclude
that, independent of the details of the scalar-field theory describing the
varying `constant', the condition for its cosmological variations to be
measured locally is almost always satisfied in physically realistic
situations. The very small differences expected to be observed between
different scales are quantified. This strengthens the proof given in our
previous paper that local experiments see global variations by dropping the
requirement of exact spherical symmetry. It provides a rigorous
justification for using terrestrial experiments and solar system
observations to constraint or detect any cosmological time variations in the
traditional `constants' of Nature in the case where non-spherical
inhomogeneities exist.

\end{abstract}

\maketitle

\section{\protect\bigskip Introduction}

Over the past few years there has been a resurgence of observational and
theoretical interest in the possibility that some of the fundamental
`constants' of Nature might be varying over cosmological timescales \cite%
{webb}. In respect of two such `constants', the fine structure constant, $%
\alpha $, and Newton's `constant' of gravitation, $G$, the idea of such
variations is not new, and was proposed by authors such as Milne \cite{milne}%
, Dirac \cite{dirac}, and Gamow \cite{gam} as a solution to some perceived
cosmological problems of the day \cite{btip}. At first, theoretical attempts
to model such variations in constants were rather crude and equations
derived under the assumption that constants like $G$ and $\alpha $ are true
constants were simply altered by writing-in an explicit time variation. This
approach was first superseded in the case of varying $G$ by the creation of
scalar-tensor theories of gravity \cite{jordan}, culminating in the standard
form of Brans and Dicke \cite{bd} in which $G$ varies through a dynamical
scalar field which conserves energy and momentum and contributes to the
curvature of spacetime by a means of a set of generalised gravitational
field equations. More recently, such self-consistent descriptions of the
spacetime variation of other constants, like $\alpha $ \cite{bek, bsm}, the
electroweak couplings \cite{ewk}, and the electron-proton mass ratio, $\mu $%
, \cite{bm} have been formulated although most observational constraints in
the literature are imposed by simply making constants into variables in
formulae derived under the assumption that are constant.

The resurgence of interest in possible time variations in $\alpha $ and $\mu
$ has been brought about by significant progress in high-precision quasar
spectroscopy. In addition to quasar spectra, we also have available a
growing number of laboratory, geochemical, and astronomical observations
with which to constrain any local changes in the values of $\ $these
constants \cite{reviews}. Studies of the variation of other constants, such
as $G$, the electron-proton mass ratio, $\mu =m_{e}/m_{pr}$, and other
standard model couplings, are confronted with an array of other data
sources. The central question which this series of papers addresses is how
to these disparate observations, made over vastly differing scales, can be
combined to give reliable constraints on the allowed global variations of $%
\alpha $ and the other constants. If $\alpha $ varies on cosmological scales
that are gravitationally unbound and participate in the Hubble expansion of
the universe, will we see any trace of this variation in a laboratory
experiment on Earth? After all, we would not expect to find the expansion of
the universe revealed by any local expansion of the Earth. In Paper I \cite%
{shawbarrow1}, we examined this question in detail for spherically symmetric
inhomogeneous universes that model the situation of a planet or a galaxy in
an expanding Friedmann-Robertson-Walker (FRW)-like universe. In this paper
we relax the strong assumption of spherical symmetry and examine the
situation of local observations in a universe that contains non-spherically
symmetric inhomogeneity. Specifically, we use the inhomogeneous metrics
found by Szekeres to describe a non-spherically symmetric universe
containing a static star or planet. As in Paper I, we are interested in
determining the difference (if any) between variations of a supposed
'constant' or associated scalar field when observed locally, on the surface
of the planet or star, and on cosmological scales.

When a `constant', $\mathbb{C}$, is made dynamical we can allow it to vary
by making it a function of a new scalar field, $\mathbb{C}\rightarrow
\mathbb{C}(\phi )$, that depends on spacetime coordinates: $\phi =\phi (\vec{%
x},t)$. It has become general practice to combine take all observational
bounds on the allowed variations of $\mathbb{C}(\phi )$. This practice
assumes implicitly that any time variation of $\mathbb{C,}$ on or near the
Earth, is comparable to any cosmological variation that it might experience,
that is to high precision
\begin{equation}
\dot{\phi}(\vec{x},t)\approx \dot{\phi}_{c}(t),  \label{wettcond}
\end{equation}%
\noindent for almost all locations $\vec{x}$, where $\phi _{c}$ is the
cosmological value of $\phi $. This assumption is always made without proof,
and there is certainly no \emph{a priori} reason why it should be valid.
Strictly, $\phi $ mediates a new or `fifth' force of Nature. If the assumed
behaviour is correct then this force is unique amongst the fundamental
forces in that its value locally reflects its cosmological variation
directly.

In this series of papers we are primarily interested in theories where the
scalar field, or `dilaton' as we shall refer to it, $\phi $, evolves
according to the conservation equation
\begin{equation*}
\square \phi =B_{,\phi }(\phi )\kappa T-V_{,\phi }(\phi ),
\end{equation*}%
where $T$ is the trace of the energy momentum tensor, $T=T_{\mu }^{\mu }$,
(with the contribution from any cosmological constant neglected). We absorb
any dilaton-to-cosmological constant coupling into the definition of $V(\phi
)$. The dilaton-to-matter coupling $B(\phi )$ and the self-interaction
potential, $V(\phi )$, are arbitrary functions of $\phi $ and units are
defined by $\kappa =8\pi G$ and $c=\hslash =1$. This covers a wide range of
theories which describe the spacetime variation of `constants' of Nature; it
includes Einstein-frame Brans-Dicke (BD) and all other, single-field,
scalar-tensor theories of gravity \cite{bd, bsm, poly, posp}. In cosmologies
that are composed of perfect fluids and a cosmological constant, it will
also contain the Bekenstein-Sandvik-Barrow-Magueijo (BSBM) theory of varying
$\alpha $, \cite{bsm}, and other single-dilaton theories which describe the
variation of standard model couplings, \cite{posp}. We considered some other
possible generalisations in \cite{shawbarrow1}. It should be noted that our
analysis and results apply equally well to any theory which involves
weakly-coupled, `light', scalar fields, and not just those that describe
variations of the standard constants of physics.

In first paper of this series, \cite{shawbarrow1}, we determined the
conditions under which condition \ref{wettcond} would hold near the surface
of a virialised over-density of matter, such as a galaxy or star, or a
planet, such as the Earth, under the assumption of spherical symmetry. We
chose to refer to this object as our `star'. In Paper I, matched asymptotic
expansions were employed to analyse the most general, \emph{%
spherically-symmetric}, dust plus cosmological constant embeddings of the
`star' into an expanding, asymptotically homogeneous and isotropic
spherically symmetric universe. We proved that, independent of the details
of the scalar-field theory describing the varying `constant', that \ref%
{wettcond} is almost always satisfied under physically realistic conditions.
The latter condition was quantified in terms of an integral over sources
that can be evaluated explicitly for any local spherical object.

In this paper we extend that analysis, and our main result, to a class of
embeddings into cosmological background universes that possess \emph{no}
Killing vectors i.e. \emph{no} symmetries. The mathematical machinery that
we use to do this is, as before, the method of matched asymptotic
expansions, employed in \cite{shawbarrow1}, where the technical machinery is
described in detail. A summary of the results obtained there can also be
found in \cite{shawbarrowlett}.

This paper is organised as follows: We shall firstly provide a very brief
summary of the method of matched asymptotic expansions used here. In section
II we will introduce the geometrical set-up that we will use. We will be
working in spacetime backgrounds of Szekeres-Szafron type \cite{szek,
szafron}.  We describe theses particular solutions of Einstein's equations
briefly in section II and then in greater detail in section III. In section
IV we extend the analysis of \cite{shawbarrow1} to include non-spherically
symmetric backgrounds of Szekeres-Szafron type.  In section V, we consider
the validity of the approximations used, and state the conditions under
which they should be expected to hold. In section VI we perform the matching
procedure (as outlined below), and extend the main result of \cite%
{shawbarrow1} to Szekeres-Szafron spacetimes. We consider possible
generalisations of our result in section VII before considering the
implications of the results in the section VIII.

We will employ the method of matched asymptotic expansions \cite{hinch,
Death}. We solve the dilaton conservation equations as an asymptotic series
in a small parameter, $\delta $, about a FRW background and the
Schwarzschild metric which surrounds our star. The deviations from these
metrics are introduced perturbatively. The former solution is called the
\emph{exterior expansion} of $\phi $, and the latter the \emph{interior
expansion} of $\phi $. The exterior expansion is found by assuming that the
length and time scales involved are of the order of some intrinsic exterior
length scale, $L_{E}$. Similarly in the interior expansion we assume all
length and time scales we be of the of $L_{I}$, the interior length scale.
Neither of the two different expansions will be valid in both regions. In
general, we define $\delta :=L_{I}/L_{E}\ll 1$. This means that in general
only a subset of our boundary conditions we will be enforceable for each
expansion, and as a result both the interior and exterior solutions will
feature unknown constants of integration. To remove this ambiguity, and
fully determine both expansions, we used the formal matching procedure. The
idea is to assume that both expansions are valid in some intermediate
region, where length scales go like $L_{int}=L_{I}^{\alpha }L_{E}^{1-\alpha }
$, for some $\alpha \in (0,1)$. Then by the uniqueness property of
asymptotic expansions, both solutions must be equal in that intermediate
region. This allows us to set the value of constants of integration, and
effectively apply \emph{all} the boundary conditions to both expansions. A
fuller discussion of this method, with examples, and its application in
general relativity is given in \cite{shawbarrow1}.

\section{Geometrical Set-Up}

We shall consider a similar geometrical set-up to that of Paper I. We assume
that the dilaton field is only weakly coupled to gravity, and so its energy
density has a negligible effect on the expansion of the background universe.
This allows us to consider the dilaton evolution on a fixed background
spacetime. We will require this background spacetime to have the same
properties as in Paper I, but with the requirement of spherical symmetry
removed:

\begin{itemize}
\item The metric is approximately Schwarzschild, with mass $m$, inside some
closed region of spacetime outside a surface at $r=R_{s}$. The metric for $%
r<R_{s}$ is left unspecified.

\item Asymptotically, the metric must approach FRW and the whole spacetime
should tend to the FRW metric in the limit $m\rightarrow 0$.

\item When the local inhomogeneous energy density of asymptotically FRW
spacetime tends to zero, the spacetime metric exterior to $r=R_{s}$ must
tend to a Schwarzschild metric with mass $m$ .
\end{itemize}

We will also limit ourselves to considering spacetimes in which the
background matter density satisfies a physically realistic equation of
state, specifically that of pressureless dust ($p=0$). We also allow for the
inclusion of a cosmological constant, $\Lambda $. The set of all
non-spherical spacetimes that satisfy these conditions is too large and
complicated for us to examine fully here; and such an analysis is beyond the
scope of this paper. We can simplify our analysis greatly, however, we
specify four further requirements:

\begin{enumerate}
\item The flow-lines of the background matter are geodesic and non-rotating.
This implies that the flow-lines are orthogonal to a family of spacelike
hypersurfaces, $S_{t}$.

\item Each of the surfaces $S_{t}$ is conformally flat.

\item The Ricci tensor for the hypersurfaces $S_{t}$, ${}^{(3)}R_{ab}$, has
two equal eigenvalues.

\item The shear tensor, as defined for the pressureless dust background, has
two equal eigenvalues.
\end{enumerate}

The last three of these conditions seem rather artificial; however, when the
deviations from spherical symmetry are in some sense `small' we might expect
them to hold as a result of the first condition. In the spherically
symmetric case, condition 1 implies conditions 2, 3 and 4. In the absence of
spherical symmetry, these conditions require the background spacetime to be
of Szekeres-Szafron type, containing pressureless matter and (possibly) a
cosmological constant. The conditions (1 - 4) combined with the background
matter being of perfect fluid type provide an invariant definition of the
Szekeres-Szafron class of metrics that is due to Szafron and Collins \cite%
{collins, kras}.

We have demanded that the `local' or interior region be approximately
Schwarzschild. The intrinsic length scale of the interior is defined by the
curvature invariant there:
\begin{equation}
L_{I}\equiv \left( \tfrac{1}{12}R_{abcd}R^{abcd}\right) ^{-1/4}=\frac{%
R_{s}^{3/2}}{\left( 2m\right) ^{1/2}}.  \label{invar}
\end{equation}%
The exterior (or cosmological) region is approximately FRW, and so its
intrinsic length scale is proportional to the inverse square root of the
local energy density: $1/\sqrt{\kappa \varepsilon +\Lambda }$, where $%
\varepsilon $ is the matter density. In accord with current astronomical
observations, we assume that this FRW region is approximately flat, and so
we set our exterior length scale appropriate for the present epoch, $t=t_{0},
$ by the inverse Hubble parameter at that time:
\begin{equation*}
L_{E}\equiv 1/H_{0}.
\end{equation*}%
We can now define a small parameter by the ratio of the interior and
exterior length scales:
\begin{equation*}
\delta =L_{I}/L_{E}.
\end{equation*}

\section{Szekeres-Szafron Backgrounds}

In 1975 Szekeres \cite{szek} solved the Einstein equations with perfect
fluid source by assuming a metric of the form:
\begin{equation*}
\mathrm{d}s^{2}=\mathrm{d}t^{2}-e^{2\alpha }\mathrm{d}r^{2}-e^{2\beta
}\left( \mathrm{d}x^{2}+\mathrm{d}y^{2}\right) ,
\end{equation*}%
with $\alpha $ and $\beta $ being functions of $(t,r,x,y)$. The coordinates
where assumed to be comoving so that the fluid-flow vector is of the form: $%
u^{\mu }=\delta _{0}^{\mu }$; This implies $p=p(t)$ and $\ $the acceleration
$\dot{u}^{\mu }=0$. Szekeres assumed a dust source with no cosmological
constant, $p=0$, although his results were later generalised to arbitrary $%
p(t)$ by Szafron \cite{szafron} and the explicit dust plus $\Lambda $
solutions were found by Barrow and Stein-Schabes \cite{JBJSS}.\emph{\ }In
general, these metrics have \emph{no} Killing symmetries \cite{bonn}.
Spherically-symmetric solutions of this type with $\alpha (r,t)$ and $\beta
(r,t)$ were, in fact, first discussed by Lemaître \cite{lem} and are usually
referred to as the Tolman-Bondi models \cite{tolbondi}; much of the analysis
of Paper I assumed a Tolman-Bondi background.

\begin{figure}[tbh]
\begin{center}
\includegraphics[scale=0.65]{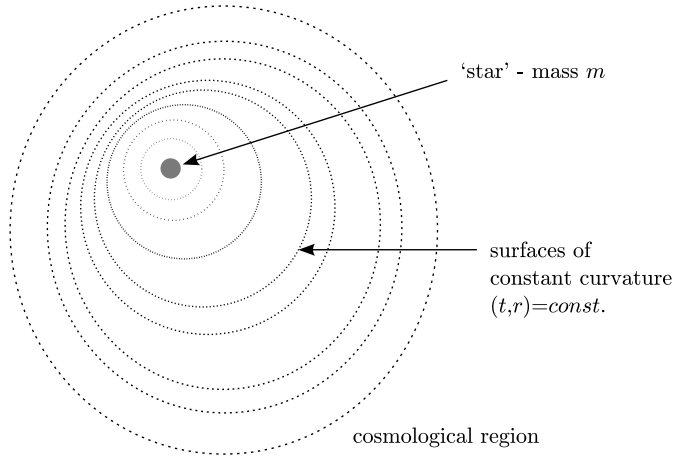}
\end{center}
\caption{The surfaces $(t,r)$ are spheres, which are concentric to leading
order in $\protect\delta $ in both the exterior and interior limits.}
\label{fig1}
\end{figure}

The Szekeres-Szafron models can be divided into two classes: $\beta _{,r}=0$
and $\beta _{,r}\neq 0$. Both classes include all FRW models in their
homogeneous and isotropic limit; however, only the latter 'quasi-spherical'
class includes the external Schwarzschild solution. Since we want to have
some part of our spacetime look Schwarzschild we will only consider the $%
\beta _{,r}\neq 0$ quasi-spherical solutions. We will also limit ourselves
to spacetimes with a cosmological constant, \cite{JBJSS}, so in effect the
total pressure is $p=-\Lambda $. These universes contain no gravitational
radiation as can be deduced from the existence of Schwarzschild as a special
case which ensures a smooth matching to Schwarzschild, which contains no
gravitational radiation. With these restrictions, $\alpha $ and $\beta $ are
given by:
\begin{eqnarray}
e^{\beta } &=&\Phi (t,r)e^{\tilde{\nu}(r,x,y)}, \\
e^{\alpha } &=&h(r)e^{-\tilde{\nu}(r,x,y)}\left( e^{\beta }\right) _{,r}, \\
e^{-\tilde{\nu}} &=&\tilde{A}(r)(x^{2}+y^{2})+2\tilde{B}_{1}(r)x+2\tilde{B}%
_{2}(r)y+\tilde{C}(r),
\end{eqnarray}%
where $\Phi (t,r)$ satisfies:
\begin{equation*}
\Phi _{,t}^{2}=-\tilde{k}(r)+2\tilde{M}(r)/\Phi +\frac{1}{3}\Lambda \Phi
^{2}.
\end{equation*}%
The functions $\tilde{A}(r)$, $\tilde{B}_{1}(r)$, $\tilde{B}_{2}(r)$, $%
\tilde{C}(r)$, $\tilde{M}(r)$, $\tilde{k}(r)$ and $\tilde{h}(r)$ are
arbitrary up to the relations:
\begin{equation*}
\tfrac{1}{4}E(r):=\tilde{A}\tilde{C}-\tilde{B}_{1}^{2}-\tilde{B}_{2}^{2}=%
\tfrac{1}{4}\left[ \tilde{h}^{-2}(r)+\tilde{k}(r)\right] .
\end{equation*}%
The surfaces $(t,r)=const$ have constant curvature $E(r)$. We will require
that the inhomogeneous region of our spacetime is localised, so that it is
by some measure finite. This implies that the surfaces of constant curvature
must be closed; we must therefore restrict ourselves to only considering
backgrounds where $E>0$. Whenever this is the case, we always can rescale
the arbitrary functions so that $E\ $can be set equal to $1$ by the
rescalings
\begin{eqnarray}
A(r):= &&\tilde{A}(r)/\sqrt{E(r)},\;B_{1}(r):=\tilde{B}_{1}(r)/\sqrt{E(r)}%
,\;B_{2}(r):=\tilde{B}_{2}(r)/\sqrt{E(r)},\;C(r):=\tilde{C}(r)/\sqrt{E(r)}%
,e^{\nu }:=\sqrt{E}e^{\tilde{\nu}}  \notag \\
k:= &&\tilde{k}(r)/E(r),\;h^{-2}:=\tilde{h}(r)^{-2}/E(r)=1-k(r),\;R(t,r):=%
\Phi (r,t)/\sqrt{E},\;M(r)=\tilde{M}(r)/E^{3/2}.  \notag
\end{eqnarray}%
These transformations can be viewed as the `gauge-fixing' of arbitrary
functions. In this gauge, $R(t,r)$ is a `physical' radial coordinate, i.e.
the surfaces $(t,r)=const$ have surface area $4\pi R^{2}$ and the metric
becomes
\begin{equation*}
\mathrm{d}s^{2}=\mathrm{d}t^{2}-\frac{\left( 1+\nu _{,R}R\right)
^{2}R_{,r}^{2}\mathrm{d}r^{2}}{1-k(r)}-R^{2}e^{2\nu }\left( \mathrm{d}x^{2}+%
\mathrm{d}y^{2}\right) ,
\end{equation*}%
\noindent where $e^{-\nu }=A(r)(x^{2}+y^{2})+2B_{1}(r)x+2B_{2}(r)y+C(r)$ and
$AC-B_{1}^{2}-B_{2}^{2}=\tfrac{1}{4}$, and $\nu _{,R}:=\nu _{,r}/R_{,r}$
and:
\begin{equation*}
R_{,t}^{2}=-k(r)+2M(r)/R+\frac{1}{3}\Lambda R^{2}.
\end{equation*}%
In this quasi-spherically symmetric subcase of the Szekeres-Szafron
spacetimes the surfaces of constant curvature, $(t,r)=const$, are 2-spheres
\cite{szek2}; however, they are not necessarily concentric. In the limit $%
\nu _{,r}\rightarrow 0$, the $(t,r)=const$ spheres becomes concentric (see
fig. \ref{fig1}). We can make one further coordinate transformation so that
the metric on the surfaces of constant curvature, $(t,r)=const$, is the
canonical metric on $S^{2}$ i.e. $\mathrm{d}\theta ^{2}+\sin ^{2}\theta
\mathrm{d}\phi ^{2}$:
\begin{eqnarray}
x\rightarrow X &=&2\left( A(r)x+B_{1}(r)\right) ,  \notag \\
y\rightarrow Y &=&2\left( A(r)y+B_{2}(r)\right) ,  \notag
\end{eqnarray}%
\noindent where $X+iY=e^{i\varphi }\cot \theta /2$. This yields
\begin{equation*}
-\nu _{,r}|_{x.y}=\frac{\lambda _{z}(X^{2}+Y^{2}-1)+2\lambda _{x}X+2\lambda
_{y}Y}{X^{2}+Y^{2}+1}=\lambda _{z}(r)\cos \theta +\lambda _{x}(r)\sin \theta
\cos \varphi +\lambda _{y}(r)\sin \theta \sin \varphi ,
\end{equation*}%
\noindent where we have defined:
\begin{eqnarray}
\lambda _{z}(r):= \frac{A^{\prime }}{A}, \qquad \lambda _{x}(r):= \left( \frac{2B_{1}}{A}\right) ^{\prime }A, \qquad \lambda _{y}(r):= \left( \frac{2B_{2}}{A}\right) ^{\prime }A.
\end{eqnarray}%
With this choice of coordinates, the local energy density of the dust
separates uniquely into a spherical symmetric part, $\varepsilon _{s}$, and
and a non-spherical part, $\varepsilon _{ns}$:
\begin{equation*}
\varepsilon =\varepsilon _{s}(t,R)+\varepsilon _{ns}(t,R,\theta ,\varphi ),
\end{equation*}%
\noindent where:
\begin{eqnarray}
\kappa \varepsilon _{s} &=&\frac{2M_{,R}}{R^{2}}, \\
\kappa \varepsilon _{ns} &=&-\frac{R\nu _{,R}}{1+\nu _{,R}R}\cdot \left(
\frac{2M}{R^{3}}\right) _{,R}.
\end{eqnarray}%
We define $M_{,R}=M_{,r}/R_{,r}$. Following the conventions of our previous
paper we write define

\begin{equation*}
M:=m+Z(r),
\end{equation*}
where $m$ is the gravitational mass of our `star'.

\subsection{Exterior Expansion}

As a result of the way that the inhomogeneity is introduced in these models,
we want the FRW limit to be `natural' , that is for the $O(3)$ orbits to
become concentric in this limit; we therefore require $\nu _{r}\sim o(1)$ as
$\delta \rightarrow 0$ in the exterior. This follows from the requirement
that the whole spacetime should become homogeneous in a smooth fashion in
the limit where the mass of our `star' vanishes: $m\rightarrow 0$. Put
another way, the introduction of our star is the only thing responsible for
making the surfaces of constant curvature non-concentric. We define, as in
the previous paper, dimensionless `radial' and time coordinates appropriate
for the exterior by

\begin{equation*}
\tau =H_{0}t,\text{ \ \ }\rho =H_{0}r.
\end{equation*}

The \emph{exterior limit} is defined by $\delta \rightarrow 0$ with $\tau $
and $\rho $ fixed. In the exterior region we find asymptotic expansions in
this limit. According to our prescription, we write

\begin{equation*}
H_{0}Z(\rho )\sim \frac{1}{2}\Omega _{m}\rho ^{3}+\delta
^{p}z_{1}(r)+o(\delta ^{p}),
\end{equation*}
and

\begin{equation*}
H_{0}^{-1}\lambda _{i}\sim \delta ^{s}l_{i}(\rho )+\mathcal{o}(\delta ^{s}).
\end{equation*}

Since $H_{0}^{-2}\left( \frac{2M}{R^{3}}\right) _{,R}\sim \mathcal{O}(\delta
^{p},\delta )$, we have that: $H_{0}^{-2}\kappa \varepsilon _{ns}\sim \mathcal{O%
}(\delta ^{p+s},\delta ^{1+s})$ whereas $H_{0}^{-2}\kappa \varepsilon _{s}\sim
\mathcal{O}(\delta ^{p},\delta )$. Thus, the non-spherical perturbation to
the energy density is always of subleading order compared to the first order
in spherical perturbation. The first-order, non-spherical, metric
perturbation appears at $\mathcal{O}(\delta ^{s})$; however, since this is
equivalent to a coordinate transform on $(r,\theta ,\varphi )$ and the
dilaton field, $\phi $, is homogeneous to leading order in the exterior,
this perturbation does not make any corrections to the dilaton conservation
equation at $\mathcal{O}(\delta ^{s})$. Thus, both at leading order, and at
next-to-leading order, both the energy density and the dilaton field will
behave in the same way as in the spherically-symmetric Tolman-Bondi case -
with the possible addition of a non-spherically symmetric vacuum
perturbation to the dilaton, $\phi $, i.e. $\phi \sim \phi _{s}+\phi
_{ns}+o(\delta ^{p})$ where $\phi _{s}$ is the spherically symmetric
solution and $\square _{FRW}\phi _{ns}=0$. As in our previous paper,
however, we are not especially interested in the exterior solution for $\phi
$ beyond zeroth order, just the effect of any background variation in $\phi $
on what is measured on the surface of a local 'star'.

\subsection{Interior Expansion}

We define dimensionless coordinates for the interior in the same way as we
did for the spherically symmetric case:

\begin{equation*}
T=L_{I}^{-1}(t-t_{0})\text{ and }\xi =R/R_{s}.
\end{equation*}

We define the \emph{interior limit} to be $\delta \rightarrow 0$ with $T$
and $\xi $ fixed, and perform out interior asymptotic expansions in this
limit. To lowest order in the interior region, we write $Z\sim \delta
^{q}R_{s}\mu _{1}$, and $\lambda _{i}:=\delta _{q^{\prime }}R_{s}^{-1}b_{i}$%
, where $i=\{x,y,z\}$. The condition that $\kappa \varepsilon >0$ everywhere
requires $q^{\prime }\geq q$ and then, to next-to-leading order, the
interior expansion of $\phi $ will be the same as it was in the
spherically-symmetric Tolman-Bondi case. We can potentially include a
non-spherical vacuum component for $\phi $; however, this will be entirely
determined by a boundary condition on $R=R_{s}$ and the need that it should
vanish for large $R$. To find the leading-order behaviour of the $\phi _{,T}$
we need to know $\phi $ at next-to-leading order. The only new case we need
to consider therefore is when $q^{\prime }=q$, i.e. $\kappa \varepsilon
_{ns}\sim \kappa \varepsilon _{s}$. In the spherically symmetric case we
considered two distinct subclasses of the Tolman-Bondi models: the flat, $k=0
$, Gautreau-Tolman-Bondi spacetimes, \cite{gautreau, kras} and the non-flat,
$k\neq 0$, Tolman-Bondi models with a simultaneous initial singularity. In
Gautreau-Tolman-Bondi models the initial singularity is non-simultaneous
from the point of view of geodesic observers. The latter class is the more
realistic, since in the former the world-lines of matter particles stream
out of the surface of our star at $R=R_{s}|_{R=R_{s}}$ i.e. $R_{,t}>0$,
whereas in the simultaneous big-bang models we can demand that matter
particles fall \emph{onto} this surface i.e. $R_{,t}|_{R=R_{s}}<0$. With
this choice, and if $R_{s}=2m$, the non-flat models properly describe the
embedding of a black hole into an expanding universe, whereas the
Gautreau-Tolman-Bondi model technically describes the embedding of a
white-hole in the same universe. In this paper we shall, therefore, only
give the results explicitly for the non-flat case -- however, we can present
a simple procedure to transform our results to the flat Gautreau case.

We define \emph{\ }

\begin{equation*}
\eta =\left( \xi ^{3/2}-3T/2\right) ^{2/3};\text{ }R_{s}\eta =r+\mathcal{O}%
(\delta ^{q},\delta ^{2/3}).
\end{equation*}%
From the exact solutions we find:
\begin{equation*}
k(\eta )=\delta ^{2/3}k_{0}\left( 1+\delta ^{q}\mu _{1}(\eta )+o\left(
\delta ^{q}\right) \right) +\mathcal{O}\left( \delta ^{5/3}\right) ,
\end{equation*}%
where
\begin{equation*}
k_{0}(\delta T)=\frac{2m}{R_{s}}\left( \frac{\pi }{H_{0}t_{0}+\delta T}%
\right) ^{2/3}.
\end{equation*}%
We can remove the $\mathcal{O}(\delta ^{2/3})$ metric perturbation by a
redefinition of the $T$ coordinate, $T\rightarrow T^{\ast }$:
\begin{equation*}
\sqrt{1-\delta ^{2/3}k_{0}}T^{\ast }=T+\int^{\xi }\frac{\sqrt{\frac{2m}{%
R_{s}\xi ^{\prime }}}\left( 1-\sqrt{1-\left( \frac{\delta ^{2/3}\pi \xi
^{\prime }}{H_{0}t_{0}+\delta T}\right) }\right) }{1-\frac{2m}{R_{s}\xi
^{\prime }}}\mathrm{d}\xi ^{\prime }.
\end{equation*}%
To leading order we see that $T\sim T^{\ast }$. The interior expansion of
the metric, for $q^{\prime }=q$, is written:
\begin{equation*}
\mathrm{d}s_{int}^{2}\sim R_{s}^{2}\left( j_{ab}^{(0)}(\xi )+\delta
^{q}j_{ab}^{(1)s}(\xi ,\chi )+\delta ^{q}j_{ab}^{(1)ns}(\xi ,\chi )+o(\delta
^{q})\right) \mathrm{d}x^{a}\mathrm{d}x^{b}+o(\delta ^{q}).
\end{equation*}%
where $j_{ab}^{(0)}$ and $j_{ab}^{(1)s}$ are given by:
\begin{eqnarray}
j_{ab}^{(0)}\mathrm{d}x^{a}\mathrm{d}x^{b} &=&\frac{R_{s}}{2m}\mathrm{d}%
T^{\ast 2}-\left( \mathrm{d}\xi +\xi ^{-1/2}\mathrm{d}T^{\ast }\right)
^{2}-\xi ^{2}\{\mathrm{d}\theta ^{2}+\sin ^{2}\theta \mathrm{d}\varphi
^{2}\},  \label{j1eq.2} \\
j_{ab}^{(1)s}\mathrm{d}x^{a}\mathrm{d}x^{b} &=&-\frac{\mu _{1}(\chi )}{\xi
^{1/2}}\mathrm{d}\xi \mathrm{d}T^{\ast }-\frac{\mu _{1}(\chi )}{\xi }\mathrm{%
d}T^{\ast 2}.
\end{eqnarray}%
These are the same as in the spherically symmetric case. The non-spherically
symmetric perturbation is given by
\begin{eqnarray}
j_{ab}^{(1)ns}\mathrm{d}x^{a}\mathrm{d}y^{b} &=&2(b_{z}\cos \theta
+b_{x}\cos \varphi \sin \theta +b_{y}\sin \varphi \sin \theta )\xi \mathrm{d}%
\eta ^{2} \\
&&-2\xi ^{2}\left[ b_{z}\sin \theta +b_{x}\cos \varphi (1-\cos \theta
)+b_{y}\sin \varphi (1-\cos \theta )\right] \mathrm{d}\theta \mathrm{d}\eta
\notag \\
&&-2\xi ^{2}(1-\cos \theta )\sin \theta (b_{x}\sin \varphi -b_{y}\cos
\varphi )\mathrm{d}\varphi \mathrm{d}\eta .  \notag
\end{eqnarray}%
\noindent The spherically symmetric part of the local energy density, $%
\kappa \varepsilon _{s}$ is the same as it was in the Tolman-Bondi cases:
\begin{equation*}
R_{s}^{2}\kappa \varepsilon _{s}=\delta ^{q}\frac{2m}{R_{s}}\frac{\mu _{1,\xi }%
}{\xi ^{3/2}\eta ^{1/2}}.
\end{equation*}%
The non-spherically symmetric part is:
\begin{equation*}
R_{s}^{2}\kappa \varepsilon _{ns}=-\delta ^{q}\frac{6m}{R_{s}}\frac{\left(
b_{z}\cos \theta +b_{x}\sin \theta \cos \phi +b_{y}\sin \theta \sin \phi
\right) }{\xi ^{3/2}\eta ^{1/2}}.
\end{equation*}%
and to ensure that the energy density is everywhere positive we need $\mu
_{,\eta }^{(1)}\geq 3b_{i}$.

\section{Extension to quasi-spherical situations}

\subsection{Boundary Conditions}

We demand the same boundary conditions as before: as the physical radius
tends to infinity, $R\rightarrow \infty $, we demand that the dilaton tends
to its homogeneous cosmological value: $\phi (R,t)\rightarrow \phi _{c}(t)$.
This can be applied to the exterior approximation. In the interior, we
demand that the dilaton-flux passing out from the surface of our `star' at $%
R=R_{s}$ is, at leading order, parametrised by:
\begin{equation}
-R_{s}^{2}\left( 1-\frac{2m}{R_{s}}\right) \left. \partial _{\xi }\phi
_{0}\right\vert _{\xi =R_{s}}=2mF\left( \bar{\phi}_{0}\right)
=\int_{0}^{R_{s}}\mathrm{d}R^{\prime }R^{\prime }{}^{2}B_{,\phi }(\phi _{0}(%
\hat{\xi}^{\prime }))\kappa \varepsilon (R^{\prime }),  \label{phiflux}
\end{equation}%
where $\bar{\phi}_{0}=\phi _{0}(R=R_{s})$. The function $F(\phi )$ can be
found by solving the dilaton field equations to leading order in the $R<R_{s}
$ region. If the interior region is a black-hole ($R_{s}=2m$) then we must
have $F(\phi )=0$; otherwise we expect $F(\phi )\sim B_{,\phi }(\phi )$.
Without considering the sub-leading order dilaton evolution inside our
`star', i.e. at $R<R_{s}$, we cannot rigorously specify any boundary
conditions beyond leading order. Despite this, we can guess at a general
boundary condition by perturbing eq. (\ref{phiflux}):
\begin{equation}
-R_{s}^{2}\left( 1-\frac{2m}{R_{s}}\right) \left. \partial _{R}\tilde{\delta}%
(\phi )\right\vert _{\xi =R_{s}}=-\left. \tilde{\delta}\left( \sqrt{-g}%
g^{RR}\right) \partial _{R}\phi _{0}\right\vert _{R=R_{s}}+2\tilde{\delta}%
(M)F\left( \bar{\phi}_{0}\right) +2mF_{,\phi }(\bar{\phi}_{0})\tilde{\delta}%
\left( \bar{\phi}_{0}\right) +\mathrm{smaller}\;\mathrm{terms},
\label{pertbdry}
\end{equation}%
where $\tilde{\delta}(X)$ is the first sub-leading order term in the
interior expansion of $X$; $M$ is the total mass contained inside $\xi <R_{s}
$ and is found by requiring the conservation of energy; and at $t=t_{0}$ we
have $M=m$. Only $\tilde{\delta}\left( \bar{\phi}_{0}\right) $ remains
unknown; however, we shall assume it to be the same order as $\tilde{\delta}%
(\phi )$ and see that this unknown term is usually suppressed by a factor of
$2m/R_{s}$ relative to the other terms in eq. (\ref{pertbdry}).

\subsection{Interior Expansion}

In the spherically symmetric case we found that $\phi \sim \phi
_{I}^{(0)}+\delta ^{q}\phi _{I}^{(1)}+o(\delta ^{q})$. In the non-spherical
case, where $q^{\prime }=q$, we relabel $\phi _{I}^{(1)}\rightarrow \phi
_{I}^{(1)s}$ and we have additional non-spherical modes:
\begin{equation*}
\phi \sim \phi _{I}^{(0)}(\xi ,T)+\delta ^{q}\phi _{I}^{(1)s}(\xi ,T)+\delta
^{q}\phi _{I}^{(1)z}(\xi ,T)\cos \theta +\delta ^{q}\phi _{I}^{(1)x}(\xi
,T)\sin \theta \cos \varphi +\delta ^{q}\phi _{I}^{(1)y}(\xi ,T)\sin \theta
\sin \varphi +o(\delta ^{q})
\end{equation*}%
where:
\begin{eqnarray}
-\frac{2m}{R_{s}}\left( \xi ^{3/2}\phi _{I,TT}^{(1)i}+\frac{3}{2}\phi
_{I,T}^{(1)i}\right)  &+&\frac{1}{\eta ^{1/2}}\left( \frac{\xi ^{5/2}}{\eta
^{1/2}}\phi _{I,\eta }^{(1)i}\right) _{,\eta }-\frac{2}{\xi ^{1/2}}\phi
_{I}^{(1)i}=\frac{6m}{R_{s}}B_{,\phi }\left( \phi _{I}^{0}\right) \frac{%
b_{i}\left( \eta \right) }{\eta ^{1/2}}  \label{nsphieqn} \\
&+&\left( \frac{2m}{R_{s}}\right) \frac{1}{\eta ^{1/2}}F\left( \bar{\phi}%
_{0}\right) \left[ \left( b_{i}(\eta )\xi \left( \frac{1+\frac{2m}{R_{s}\xi }%
}{1-\frac{2m}{R_{s}\xi }}\right) \right) _{,\eta }-2b_{i}(\eta )\right] .
\notag
\end{eqnarray}%
We can solve this order by order in $2m/R_{s},$ and to lowest order we find:
\begin{eqnarray}
\phi _{I}^{(1)i} &\sim &\frac{2m}{R_{s}}B_{,\phi }\left( \phi
_{I}^{0}\right) \xi \int^{\eta }\mathrm{d}\eta ^{\prime }\frac{b_{i}(\eta
^{\prime })}{\xi ^{^{\prime }2}}-\frac{2m}{R_{s}}B_{,\phi }\left( \phi
_{I}^{0}\right) \frac{1}{\xi ^{2}}\int_{\xi =1}^{\eta }\mathrm{d}\eta
^{\prime }\xi ^{\prime }b_{i}(\eta ^{\prime }) \\
&+&\frac{2m}{R_{s}}F\left( \bar{\phi}_{0}\right) \frac{1}{\xi ^{2}}\int_{\xi
=1}^{\eta }\mathrm{d}\eta ^{\prime }b_{i}(\eta ^{\prime })\xi ^{\prime }+%
\frac{C_{i}}{\xi ^{2}}+D_{i}\xi +\mathcal{O}((2m/R_{s})^{2})  \notag
\end{eqnarray}%
Since we are interested in finding when and where the local time variation
of $\phi $ deviates from its cosmological value, we are chiefly concerned
with the case $q\leq 1$. The matching condition then requires that we fix $%
D_{i}$ so that in the intermediate limit we have $\phi _{I}^{(1)i}\sim \xi
^{n}$ with $n<1$. The value of $C_{i}$ should be set by a boundary condition
on $R=R_{s}$. We cannot specify $C_{i}$ exactly without further information
about the interior of our `star' in $R<R_{s}$. If we assume that the
prescription for the sub-leading order boundary condition given above is
correct then we find:
\begin{eqnarray}
\partial _{\xi }\phi _{I}^{(1)i}|_{\xi =1} &\sim &\frac{2m}{R_{s}}\left.
\frac{b_{i}}{\eta ^{1/2}}\right\vert _{\xi =1}F\left( \bar{\phi}_{0}\right) +%
\mathcal{O}((2m/R_{s})^{2})  \notag \\
\Rightarrow C_{i} &=&\frac{m}{R_{s}}B_{,\phi }\left( \phi _{I}^{0}\right)
\int^{\xi =1}\mathrm{d}\eta ^{\prime }\frac{b_{i}(\eta ^{\prime })}{\xi
^{^{\prime }2}}+\tfrac{1}{2}D  \notag
\end{eqnarray}%
From now onwards we set $C_{i}=0,$ for simplicity; even when this is not
correct we do not expect the magnitude of $C_{i}$ or $C_{i,T}$ to be larger
than any of the other terms in $\phi _{I}^{(1)i}$ or $\phi _{I,T}^{(1)i}$,
respectively. The time-derivative of $\phi _{I}^{(1)i}$ for fixed $R$ is:
\begin{equation}
\phi _{I,T}^{(1)i}\sim \frac{4m}{R_{s}}B_{,\phi }\left( \phi _{I}^{0}\right)
\xi \int^{\eta }\mathrm{d}\xi ^{\prime }\frac{b_{i}(\eta ^{\prime })}{\xi
^{^{\prime }5/2}}+\frac{2m}{R_{s}}B_{,\phi }\left( \phi _{I}^{0}\right)
\frac{1}{\xi ^{2}}\int_{\xi =1}^{\eta }\mathrm{d}\xi ^{\prime }\frac{%
b_{i}(\eta ^{\prime })}{\xi ^{1/2}}-\frac{2m}{R_{s}}F\left( \bar{\phi}%
_{0}\right) \frac{1}{\xi ^{2}}\int_{\xi =1}^{\eta }\mathrm{d}\eta ^{\prime }%
\frac{b_{i}(\eta ^{\prime })}{\xi ^{1/2}}+D_{,T}\xi   \label{nsphitev}
\end{equation}%
In the next section we shall discuss what we require of the $b_{i}$ for the
matching procedure to be valid. In section VI we will then use the matching
conditions to find $D$ and $D_{,T}$.

We could also relax the requirement that the leading-order mode in $\phi $
be spherically symmetric. At next-to-leading order these new modes would
generate extra terms in $\phi _{I}^{(1)}$. In general, an $l$-pole at
leading order becomes an $l+1$-pole at next-to-leading order. The magnitude
of the extra time-dependence that is picked up is, however, the same each
time. Hence, we restrict ourselves by taking the leading-order mode to be
spherically symmetric for the time being. Note also that we can pass from
the simultaneous big-bang case, to the spatially flat, `Gautreau', case by
setting $k=0$ and making the transform $\eta \rightarrow \chi =\left( \xi
^{3/2}+3T/2\right) ^{2/3}$. This will also mean that $\phi _{I,T}\rightarrow
-\phi _{I,T}$.

\section{Validity of Approximations}

All of the conditions found in Paper I for the matching of the spherically
symmetric parts of $\phi $ to be possible still apply here. However, we must
now satisfy some extra conditions that come from the requirement that the
non-spherical parts should also be matchable.

We assume that $b_{i}\left( \eta \right) \propto \eta ^{d_{i}}$ as $\eta
\rightarrow \infty $ for some $d_{i}>0$. At order $\delta ^{q}$, the growing
mode in the non-spherically symmetric part of the interior approximation
will then grow like $\delta ^{q}\eta ^{d_{i}+1}/\xi $. In the intermediate,
or matching, region we have that $\eta ,\xi \sim \delta ^{-\alpha }$ for
some $\alpha \in (0,1)$. We require $\phi _{I}$ to have a valid asymptotic
expansion this region. This implies that there exists some $\alpha \in (0,1)$
such that, for each $i$, we have $\alpha -q/d_{i}>0$.

In the exterior we shall write $H_{0}^{-1}\lambda _{i}\sim \delta
^{p_{i}^{\prime }}l_{i}(\rho )$, where $p_{i}^{\prime }>0$ comes from the
requirement that the 2-spheres of constant curvature become concentric in
the exterior limit. As $\rho \rightarrow 0$ we assume that $l_{i}(\rho
)\propto \rho ^{-f_{i}}$. We previously stated that $Z\sim \frac{1}{2}\Omega
_{m}\rho ^{3}+\delta ^{p}z_{1}+o(\delta ^{p})$ in the exterior. We assume
that as $\rho \rightarrow 0$, we have $z_{1}\propto \rho ^{-m}$. Although we
did not explicitly consider the exterior expansion of $\phi $ we can now
examine the behaviour of the leading-order non-spherically symmetric mode in
the intermediate limit of that exterior expansion. We noted above that there
will be no $\mathcal{O}(\delta ^{p_{i}^{\prime }})$ correction resulting
from the $l_{i}$. The leading-order mode will therefore either go like $%
\max_{i}\left( \delta ^{p+p_{i}^{\prime }}z_{1}(\rho )l_{i}(\rho )\right) $
if $p<1$ or $\max_{i}\left( \delta ^{1+p_{i}^{\prime }}(\rho )l_{i}(\rho
)\right) $ otherwise, and $\rho \sim \mathcal{O}(\delta ^{1-\alpha })$ in
the intermediate region. For the exterior expansion to be valid in the
intermediate region we therefore require
\begin{eqnarray}
\max_{i}(p_{i}^{\prime } &+&(1-\alpha )(f_{i}+m))>-p\;\mathrm{if}\;p\leq 1,
\notag \\
\max_{i}(p_{i}^{\prime } &+&(1-\alpha )f_{i})>-1\;\mathrm{if}\;p\geq 1.
\notag
\end{eqnarray}

These conditions on $\alpha $ are equivalent to the following: there exists $%
\alpha $ such that the interior expansion of $R^{2}\kappa \varepsilon _{ns}$ is
$o(1)$ as $\delta \rightarrow 0$ for all $0<\alpha ^{\prime }<\alpha $ where
$\xi ,T\sim \mathcal{O}(\delta ^{-\alpha })$, and the exterior expansion of $%
R^{2}\kappa \varepsilon _{ns}$ is also $o(1)$ as $\delta \rightarrow 0$ for all
$0<\alpha ^{\prime \prime }<\alpha $ where $\rho ,\tau -\tau _{0}\sim
\mathcal{O}(\delta ^{1-\alpha })$. This suggests that the condition for the
matching procedure to work, as far as the spherically non-symmetric modes
are concerned, is simply that
\begin{equation*}
R^{2}\kappa \varepsilon _{ns}\ll 1\;\mathrm{everywhere.}
\end{equation*}%
We can also rephrase and generalise the conditions for the matching
procedure to be possible w.r.t. the spherically symmetric modes (as found in
\cite{shawbarrow1}) in a similar fashion: for all $\alpha \in (0,1)$, and
keeping $L_{I}^{\alpha }L_{E}^{1-\alpha }(t-t_{0}),L_{I}^{\alpha
}L_{E}^{1-\alpha }R$ fixed, we have $\lim_{\delta \rightarrow 0}{R^{2}\kappa
\Delta \varepsilon _{s}}=o(1)$ and $\lim_{\delta \rightarrow 0}{2(m+Z)/R}=o(1)$%
. We can combine our two conditions by simply replacing $\Delta \varepsilon _{s}
$ by $\Delta \varepsilon $ in the above expression. Strictly speaking, since $%
\alpha \in (0,1)$ (as opposed to $[0,1)$, $(0,1]$ or $[0,1]$) we can also
replace $\Delta \varepsilon $ by just $\varepsilon $ since $R^{2}\kappa \varepsilon
_{FRW}$ is small everywhere outside the exterior region. For Szekeres
backgrounds the first of these conditions implies the second everywhere
outside the interior region. Therefore, the matching procedure is certainly
possible to zeroth order, if:
\begin{equation*}
\forall \alpha \in (0,1):\;\;\lim_{\delta \rightarrow 0}\left( R^{2}\kappa
\varepsilon (R,t)\right) =o(1)\;\mathrm{and}\;\;\lim_{\delta \rightarrow
0}\left( M(R,t)/R\right) =o(1)\;\;\mathrm{with}\;\;\{L_{I}^{\alpha
}L_{E}^{1-\alpha }(t-t_{0}),L_{I}^{\alpha }L_{E}^{1-\alpha }R\}\;\mathrm{%
fixed},
\end{equation*}%
\noindent where $M(R,t)$ is the gravitational mass inside the surface $%
(t,R)=const$. Equivalently, in \emph{any} intermediate region the background
spacetime is asymptotically Minkowski as $\delta \rightarrow 0$: everywhere
which is not in either the interior or exterior regions can be considered to
be a weak-field perturbation of Minkowski spacetime. The power of our method
is that we do \emph{not} require this to be true of the interior and
exterior regions. So long as this condition holds in the intermediate
region, we can match the zeroth-order approximations in some region and find
the circumstances under which condition (\ref{wettcond}) holds by comparing
the relative sizes of the derivatives $\phi _{c,t}$ and $\phi _{I,t}^{(1)}$.

\section{Matching}

We rewrite the expression for the $\phi _{I}^{(1)i}$ in terms of the
non-spherical part of local density:
\begin{eqnarray}
\delta ^{q}\phi _{I}^{(1)ns} &=&\delta ^{q}\left( \phi _{I,t}^{(1)z}\cos
\theta +\phi _{I,t}^{(1)x}\sin \theta \cos \varphi +\phi _{I,t}^{(1)y}\sin
\theta \sin \varphi \right)   \notag \\
&\sim &-\frac{1}{3}B_{,\phi }\left( \phi _{I}^{0}\right) R\int^{r}\mathrm{d}%
r^{\prime }R_{,r}\kappa \varepsilon _{ns}(r^{\prime },t)-\frac{R}{R_{s}}\hat{D}%
(T,\theta ,\phi )  \notag \\
&&-\frac{1}{3}B_{,\phi }\left( \phi _{I}^{0}\right) \frac{1}{R^{2}}%
\int_{R=R_{s}}^{r}\mathrm{d}r^{\prime }R_{,r}R^{3}\kappa \varepsilon
_{ns}(r^{\prime },t)-\frac{1}{3}F\left( \bar{\phi}_{0}\right) \frac{1}{R^{2}}%
\int_{R=R_{s}}^{r}\mathrm{d}r^{\prime }R_{,r}R^{3}\kappa \varepsilon
_{ns}(r^{\prime },t)  \notag
\end{eqnarray}%
\noindent where $\hat{D}(T,\theta ,\phi ):=D_{z}\cos \theta +D_{x}\sin
\theta \cos \varphi +D_{y}\sin \theta \sin \varphi $. By examining the
dilaton equations of motion in the FRW region, we can see there is a
component of the leading-order $(\theta ,\varphi )$-dependent term in the
exterior expansion or $\phi $ behaves like
\begin{equation*}
-\frac{1}{3}B_{,\phi }(\phi _{c})R\int_{\infty }^{r}\mathrm{d}r^{\prime
}R_{,r}\kappa \varepsilon _{ns}(r^{\prime },t)
\end{equation*}%
\noindent for $R\ll H_{0}^{-1}$ and $t$ fixed. Therefore matching requires
that we choose $\hat{D}$ such that
\begin{eqnarray}
\delta ^{q}\phi _{I}^{(1)ns} &=&\delta ^{q}\left( \phi _{I,t}^{(1)z}\cos
\theta +\phi _{I,t}^{(1)x}\sin \theta \cos \varphi +\phi _{I,t}^{(1)y}\sin
\theta \sin \varphi \right)  \\
&\sim &-\frac{1}{3}B_{,\phi }\left( \phi _{I}^{0}\right) R\int_{\infty }^{r}%
\mathrm{d}r^{\prime }R_{,r}\kappa \varepsilon _{ns}(r^{\prime },t)+\frac{1}{3}%
B_{,\phi }\left( \phi _{I}^{0}\right) \frac{1}{R^{2}}\int_{R=R_{s}}^{r}%
\mathrm{d}r^{\prime }R_{,r}R^{3}\kappa \varepsilon _{ns}(r^{\prime },t) \\
&&-\frac{1}{3}F\left( \bar{\phi}_{0}\right) \frac{1}{R^{2}}\int_{R=R_{s}}^{r}%
\mathrm{d}r^{\prime }R_{,r}R^{3}\kappa \varepsilon _{ns}(r^{\prime },t).  \notag
\end{eqnarray}%
The interior expansion is now fully specified to order $\mathcal{O}(\delta
^{p})$. We are interested in the behaviour of $\phi _{I,t}$ and we find
\begin{eqnarray}
\delta ^{q}\phi _{I,t}^{(1)ns} &\sim &\frac{2}{3}B_{,\phi }\left( \phi
_{I}^{0}\right) R\int_{\infty }^{r}\mathrm{d}r^{\prime }R_{,r}R_{,t}\frac{%
\kappa \varepsilon _{ns}(r^{\prime },t)}{R}+\frac{1}{3}B_{,\phi }\left( \phi
_{I}^{0}\right) \frac{1}{R^{2}}\int_{R=R_{s}}^{r}\mathrm{d}r^{\prime
}R_{,r}R_{,t}R^{2}\kappa \varepsilon _{ns}(r^{\prime },t) \\
&&-\frac{1}{3}F\left( \bar{\phi}_{0}\right) \frac{1}{R^{2}}\int_{R=R_{s}}^{r}%
\mathrm{d}r^{\prime }R_{,r}R_{,t}R^{3}\kappa \varepsilon _{ns}(r^{\prime },t)+%
\frac{1}{3}F\left( \bar{\phi}_{0}\right) RR_{,t}\kappa \varepsilon _{ns}(r,t).
\notag
\end{eqnarray}%
This expression is valid whenever $R_{s}\gg 2m$, and the requirements for
matching are satisfied. In these cases we expect $F\left( \bar{\phi}%
_{0}\right) \approx B_{,\phi }\left( \phi _{I}^{0}\right) +\mathcal{O}%
(2m/R_{s})$; so, approximately, we have
\begin{equation*}
\delta ^{q}\phi _{I,t}^{(1)ns}\sim \frac{2}{3}B_{,\phi }\left( \phi
_{c}\right) R\int_{\infty }^{r}\mathrm{d}r^{\prime }R_{,r}R_{,t}\frac{\kappa
\varepsilon _{ns}(r^{\prime },t)}{R}+\frac{1}{3}B_{,\phi }\left( \phi
_{c}\right) RR_{,t}\kappa \varepsilon _{ns}(r,t).
\end{equation*}%
In the case, where $R_{s}=2m,$ and our `star' is actually a black-hole, we
require $F\left( \bar{\phi}_{0}\right) $ to ensure that the $\phi $ is
well-defined as $R\rightarrow 2m$. Even so, in this case, equation (\ref%
{nsphitev}) will not be strictly valid, since it was derived under the
assumption of $R_{s}\gg 2m$. By inspection of the dilaton evolution equation
in the interior, eq. (\ref{nsphieqn}), however, we expect that $\delta
^{q}\phi _{I,t}^{(1)ns}$ near the black-hole horizon to be of similar
magnitude to the RHS of eq. (\ref{nsphitev}).

Combining the results of this paper with those for the spherically symmetric
case we find:
\begin{equation*}
\phi _{I,t}-\phi _{c,t}\sim B_{,\phi }\left( \phi _{c}\right) \int_{\infty
}^{r}\mathrm{d}r^{\prime }R_{,r}R_{,t}\kappa \Delta \varepsilon _{s}(r^{\prime
},t)+\frac{2}{3}B_{,\phi }\left( \phi _{c}\right) R\int_{\infty }^{r}\mathrm{%
d}r^{\prime }R_{,r}R_{,t}\frac{\kappa \varepsilon _{ns}(r^{\prime },t)}{R}+%
\frac{1}{3}B_{,\phi }\left( \phi _{c}\right) RR_{,t}\kappa \varepsilon
_{ns}(r,t).
\end{equation*}%
We require that $|(\phi _{I,t}-\phi _{c,t})/\phi _{c,t}|\ll 1$ for \ref%
{wettcond} to hold and so ensure that local observations will detect
variations of $\phi $ occurring on cosmological scales.

\section{Generalisation: a conjecture}

So far, we have found an analytic approximation to the values of $\phi $ and
$\phi _{c,t}$ in the interior. More succinctly (although less explicitly) we
can say that, to leading order in $\delta $, the values of $\phi $, $\phi
_{,t}$ and $\phi _{,r}$ can all be found everywhere outside the exterior
region from the approximation:
\begin{equation}
\phi \approx \phi _{hom}(t)+\phi _{l}(\vec{x},t),  \label{phicon}
\end{equation}%
where $\phi _{l}$ is the solution to:
\begin{equation*}
\square _{sch}\phi _{l}=B_{,\phi }\kappa \Delta \varepsilon
\end{equation*}%
with $\Delta \varepsilon =\varepsilon (\vec{x},t)-\varepsilon _{c}(t)$, $\square
_{sch}$ is the wave operator in a Schwarzschild background, and $t$ is the
proper time of a comoving observer. This is solved w.r.t. the boundary
conditions $\phi _{l}\rightarrow 0$ as $R\rightarrow \infty $ (where $R=0$
is the centre of our `star') and the flux out of the `star' is as given by
equations (\ref{phiflux}) and (\ref{pertbdry}). The homogeneous term is

\begin{equation*}
\phi _{hom}(t)=\phi _{c}(t+\Delta t(\vec{x},x))
\end{equation*}%
where the \emph{lag}, $\Delta t(\vec{x},t)$, is defined by:
\begin{equation*}
\vec{\nabla}^{2}\Delta t-\vec{v}^{\ast }\cdot \vec{\nabla}(\vec{v}^{\ast
}\cdot \vec{\nabla}\Delta t)-\vec{v}^{\ast }\cdot \vec{\nabla}\Delta t(\vec{%
\nabla}\cdot \vec{v}^{\ast })=-\vec{\nabla}\cdot \Delta \vec{v},
\end{equation*}%
with $\nabla _{i}=\partial _{i}$, $i=\{1,2,3\}$, and $\Delta v=\vec{v}-H\vec{%
x}$, where $\vec{v}$ is the velocity of the dust particles relative to $%
R=\Vert \vec{x}\Vert =0$. The velocity $\vec{v}^{\ast }$ has the following
properties: $\vec{v}^{\ast }=\vec{v}$ in some region that includes all the
interior and excludes all of the exterior; $\vec{v}^{\ast }=\Delta \vec{v}$
everywhere else. In a general sense, the interior and exterior are two
disjoint regions of total spacetimes where general-relativistic effects are
non-negligible at leading order (e.g. such as when $\Vert \vec{v}\Vert
\approx 1$). The interior region should be closed, and in the exterior
region $\Vert \Delta \vec{v}\Vert $ is small. So, $\vec{v}^{\ast }$ should
be defined in such a way that it respects all the symmetries of the
spacetime and so that $\Vert \vec{v}^{ast}\Vert \ll 1$ everywhere outside
the interior region. This is required to ensure that $\Delta t$, as defined
above, is finite. It can be seen to come out of the matching procedure. When
the background spacetime satisfies the conditions given below, the precise
way in which $\vec{v}^{\ast }$ is defined does not effect the leading order
behavior of $\Delta t$. For boundary conditions, we must require the flux
out of $\Delta t$ out of the `star' to vanish, and require $\Delta
t\rightarrow 0$ as $\Delta v\rightarrow 0$, i.e. as $R\rightarrow \infty $.
This is the natural generalisation of what has been seen in the
Szekeres-Szafron backgrounds $\vec{v}^{i}=R_{,t}(R,t)\delta _{R}^{i}$. In
these cases the equation is just an ordinary differential equation in $R$
with solution:
\begin{equation*}
\Delta t=\int_{R}^{A}\mathrm{d}R^{\prime }\frac{(R_{,t}(R^{\prime
},t)-HR^{\prime }+R_{,t}(R_{s},t)+HR_{s})}{1-R_{,t}^{2}(R^{\prime },t)}%
+\int_{A}^{\infty }\mathrm{d}R^{\prime }\frac{(R_{,t}(R^{\prime
},t)-HR^{\prime }+R_{,t}(R_{s},t)+HR_{s})}{1-(R_{,t}(R^{\prime },t)-HR)^{2}},
\end{equation*}%
where $A$ is some arbitrary value of $A$ in the intermediate region, and
each $A$ represents a particular choice of definition for $\vec{v}^{\ast }$.
This expression is only valid to leading order in the interior and
intermediate regions. To this order all choices for $A$ are equivalent. Near
$R=R_{s}$, to leading order in $\delta =L_{I}/L_{E}$, this ensures that $%
\mathrm{d}(t+\Delta t)\sim \mathrm{d}v$, where $v=t_{sch}+R+2m\ln (R/2m-1)$
is the advanced time coordinate and $t_{sch}$ is the standard,
curvature-defined, Schwarzschild time-coordinate. The solution for $\phi
_{hom}$ is then, to leading order in $\delta $, just the particular one
given by Jacobson in \cite{jacobson}. We have assumed that the
generalisations of the Szekeres-Szafron result for $\phi $ hold. We have
only proved that this assumption holds for the subset of Szekeres-Szafron
spacetimes for which the matching procedure works. Nonetheless, based on
this analysis, we conjecture that \ref{phicon} provides a good numerical
approximation to the value of $\phi $, and by differentiating once, to $\phi
_{,t}$ and $\partial _{i}\phi $, $i=\{1,2,3\}$, near the surface of our
`star', for any dust plus $\Lambda $ spacetime that can be everywhere
considered to be a weak-field perturbation of either Schwarzschild,
Minkowski, or FRW spacetime; that is,
\begin{equation*}
R^{2}\kappa \Delta \varepsilon (R,t)\ll 1,\qquad 2(M(R,t)-m)/R\ll 1
\end{equation*}%
where $M(R,t)$ is the gravitational mass contained inside the surface $%
(R,t)=const$. One could seek to motivate our conjecture as some sort of
analytical continuation from the Szekeres-Szafron spacetimes to more general
backgrounds, but such arguments would, we believe, be hard to frame in any
rigorous context and are beyond the scope of the analysis in this paper.

\section{Discussion}

In this paper we have extended the analysis of \cite{shawbarrow1} to include
a class of dust-filled spacetimes without any symmetries provided by the
Szekeres-Szafron metrics. Again, we have used the method of matched
asymptotic expansions to link the evolution of the dilaton field, $\phi $,
in an approximately Schwarzschild region of spacetime to its evolution in
the cosmological background. By these methods, we have provided a rigorous
construction of what has been simply assumed about the matching procedures
in earlier studies \cite{early}. We have also analysed, more fully, the
conditions that we need the background spacetime to satisfy for the matching
procedure to be valid, and we have interpreted these conditions in terms of
their requirements on the local energy density. Finally, we have conjectured
a generalisation of our result to more general spacetime backgrounds than
those considered here.

By combining the results found here with those of the previous paper, we
conclude that, in the class of quasi-spherical Szekeres spacetimes in which
the matching procedure is valid, the local time variation of the dilaton
field will track its cosmological value whenever:
\begin{equation}
\left\vert \frac{B_{,\phi }\left( \phi _{c}\right) \int_{\infty }^{r}\mathrm{%
d}r^{\prime }R_{,r}\kappa \Delta (R_{,t}\varepsilon _{s}(r^{\prime },t))+\frac{2%
}{3}B_{,\phi }\left( \phi _{c}\right) R\int_{\infty }^{r}\mathrm{d}r^{\prime
}R_{,r}R_{,t}\frac{\kappa \varepsilon _{ns}(r^{\prime },t)}{R}+\frac{1}{3}%
B_{,\phi }\left( \phi _{c}\right) RR_{,t}\kappa \varepsilon _{ns}(r,t)}{\dot{%
\phi}_{c}}\right\vert \ll 1.  \label{condition1}
\end{equation}%
When the cosmological evolution of $\phi $ is dominated by its matter
coupling: $\dot{\phi}_{c}\sim \mathcal{O}(B_{,\phi }H_{0}^{-1}\kappa
\varepsilon _{c}),$ this condition is equivalent to:
\begin{equation*}
\left\vert H_{0}\int_{\infty }^{r}\mathrm{d}r^{\prime }R_{,r}\frac{\Delta
(R_{,t}\varepsilon _{s}(r^{\prime },t))}{\varepsilon _{c}(t)}+\frac{2}{3}%
H_{0}R\int_{\infty }^{r}\mathrm{d}r^{\prime }R_{,r}R_{,t}R^{-1}\frac{%
\varepsilon _{ns}(r^{\prime },t)}{\varepsilon _{c}(t)}+\frac{1}{3}H_{0}R_{,t}\frac{%
\varepsilon _{ns}(r,t)}{\varepsilon _{c}(t)}\right\vert \ll 1.
\end{equation*}

In the other extreme, when the potential term dominates the cosmic dilaton
evolution, the left-hand side of the above condition is further suppressed
by a factor of $B_{,\phi }(\phi _{c})/V_{,\phi }(\phi _{c})\ll 1$. As in our
previous paper, \cite{shawbarrow1}, we can see that for a given evolution of
the background matter density, condition (\ref{wettcond}) is more likely to
hold (or will hold more strongly) when $\left\vert B_{,\phi }(\phi
_{c})\kappa \varepsilon _{c}/V_{,\phi }(\phi _{c})\right\vert \ll 1$. We
reiterate our previous statement that: \emph{domination by the potential
term in the cosmic evolution of the dilaton has a homogenising effect on the
time variation of} $\phi $.

The non-spherically symmetric parts of energy density enter into the
expression differently. The magnitude of the terms on the left-hand side of
eq. (\ref{condition1}) is, as in the spherically symmetric case, still $%
\left\langle H_{0}R\Delta R_{,t}\varepsilon /\varepsilon \right\rangle (R,t)$
where $\left\langle \cdot \right\rangle (R,t)$ represents some `average'
over the region outside the surface $(R,t)=const$. We should note that,
given the condition on $\kappa \varepsilon $ that has been required for
matching, the leading-order contribution to $\kappa \varepsilon _{ns}$ is
everywhere of dipole form and this is responsible for the special form of
the average over the non-spherically symmetric terms. We can also see that,
as a result of form of eqn. (\ref{condition1}), peaks in $\kappa \varepsilon
_{ns}$ that occur outside of the interior region will, in the interior,
produce a weaker contribution to the left-hand side of eqn. (\ref{condition1}%
) than a peak of similar amplitude in a spherically symmetric energy density
$\kappa \varepsilon _{s}$. This behaviour would continue if we were also to
account for higher multipole terms in $\kappa \varepsilon _{ns}$. The higher
the multipole, the more `massive' the mode, and the faster it dissipates.

If we are interested in finding a sufficient condition (as opposed to a
necessary and sufficient one) for (\ref{wettcond}) to hold locally, then in
most circumstances we will be justified in averaging over the
non-spherically symmetric modes in the same way as we average over the
spherically symmetric ones. In most cases, this will over-estimate rather
than under-estimate the magnitude of the left-hand side of our condition, (%
\ref{condition1}). This reasoning leads us to the statement that for $\dot{%
\phi}(\mathbf{x},t)\approx \dot{\phi}_{c}(t)$ to hold locally it is
sufficient that:
\begin{equation}
\mathcal{I}:=\int_{\gamma (R)}\mathrm{d}lH_{0}R^{\prime }\frac{\Delta
(v\varepsilon )}{\varepsilon _{c}}\ll 1  \label{suff}
\end{equation}%
\noindent where $\mathrm{d}l:=\mathrm{d}rR_{,r}$, $v=R_{,t}$ is the velocity
of the dust particles, $\lim_{R\rightarrow \infty }v=H_{0}R$. We make the
same generalisation that we did in Paper I by taking $\gamma (R)$ to run
from $R$ to spatial infinity along a past, radially-directed light-ray. In
this way, we incorporate the limitations imposed by causality. We should
also assume that the above expression includes some sort of average over
angular directions; to be safe we could replace $\varepsilon $ by its maximum
value for fixed $R$ and $t$. This sufficient condition, (\ref{suff}), is
precisely the generalised condition proposed in our first paper on this
issue. The inclusion of deviations from spherical symmetry, therefore, has
little effect of the qualitative nature of the conclusions that were found
in \cite{shawbarrow1}. If anything, we have seen that the non-spherical
modes dissipate faster and, as a result, will produce smaller than otherwise
expected deviations in the local time derivative of $\phi $ from the
cosmological ones.

On Earth we should expect, as before, that the leading-order deviation of $%
\dot{\phi}$ from $\dot{\phi}_{c}$ is produced by the galaxy cluster in which
we sit, and that for a dilaton evolution that is dominated by its coupling
to matter, this effect gives $\mathcal{I}\approx 6\times 10^{-3}\Omega
_{m}^{-1}h\ll 1$, where $\Omega _{m}\approx 0.27$ and $h\approx 0.71$. If
the cosmic dilaton evolution is potential dominated then $\mathcal{I}$ is
even smaller. We conclude, as before, that irrespective of the value of the
dilaton-to-matter coupling, and what dominates the cosmic dilaton evolution,
that
\begin{equation*}
\dot{\phi}(\mathbf{x},t)\approx \dot{\phi}_{c}(t)
\end{equation*}%
will hold in the solar system in general, and on Earth in particular, to a
precision determined by our calculable constant $\mathcal{I}$. We also
conclude, as before, that whenever $\mathcal{I}\ll 1$ near the horizon of a
black hole, there will be no significant gravitational memory effect for
physically reasonable values of the parameters \cite{memory, jacobson}.

Our result relies on one major assumption: the physically realistic
condition that the scalar field should be weakly coupled to matter and
gravity -- in effect, the variations of 'constants' on large scales must
occur more slowly than the universe is expanding and so their dynamics have
a negligible back-reaction on the cosmological background metric. In this
paper we have removed the previous condition of spherical symmetry at least
in as far as the spacetime background is well described by Szekeres-Szafron
solution. We have therefore extended the domain of applicability our general
proof: that \emph{terrestrial} and \emph{solar system} based observations
can legitimately be used to constrain the \emph{cosmological} time variation
of supposed `constants' of Nature and other light scalar fields.

\begin{acknowledgments}
\bigskip We thank Tim Clifton and Peter D'Eath for discussions. D. Shaw is
supported by a PPARC studentship.
\end{acknowledgments}

\end{document}